\documentclass[amsmath, aps, prb, twocolumn, showpacs]{revtex4}
\usepackage{graphicx}

\begin{document}
  \title{A density-functional study of charge doping in WO$_3$}

  \author{Andrew D. Walkingshaw}
  \author{Nicola A. Spaldin}
  \altaffiliation{Permanent address: Materials Department, University of
    California, Santa Barbara, CA 93106} 
  \author{Emilio Artacho} 
  \affiliation{Department of Earth Sciences, University of Cambridge,
    Downing Street, Cambridge CB2 3EQ, UK}

  \date{\today}

  \begin{abstract}

    The addition of electron donors to the vacant A site of
    defect-perovskite structure tungsten trioxide causes a series of
    structural and chemical phase transitions; for instance, in the
    well-known case of the sodium tungsten bronzes (Na$_x$WO$_3$).
    Here we calculate the effect of the addition of electronic charge
    to WO$_3$ without the complication of also including sodium
    ions. Our density functional theory method enables isolation of
    electronic effects from the additional size, chemical, and
    disorder effects present in experimental samples. Our calculated
    low-temperature phase diagram between $x$=0 and $x$=1 moves from
    the initial low-temperature monoclinic phase through a second
    (centered) monoclinic phase, an orthorhombic phase, a tetragonal
    antiferroelectric, and an aristotypic cubic phase, in broad
    agreement with the experimentally-observed transformations in
    Na$_x$WO$_3$. Our work confirms that the observed structural
    transformations are driven primarily by electronic factors. We
    find that the dominant electronic effect is the covalent
    interaction between the tungsten $5d$ and oxygen $2p$ orbitals.
    
  \end{abstract}

  \maketitle

  \section{Introduction}

  Tungsten trioxide, WO$_3$, is a technologically-significant ceramic:
  it is electrochromic\cite{elc1}, and the possibility of ion
  intercalation and deintercalation gives rise to several potential
  applications in devices (for instance, as a cathode of rechargable
  batteries\cite{LiWO3batteries}). In particular, it is often used in
  optical applications due to the fact that the color can be
  changed by doping with electrons.  (This was first observed in 1815
  by Berzelius\cite{Berzelius} in H$_x$WO$_3$). Bulk stoichiometric
  WO$_3$ is yellow-green in hue, but the sodium-doped tungsten
  bronzes, Na$_x$WO$_3$, exhibit most colors of the visible spectrum
  on varying Na concentration\cite{Brown54}. In addition, the optical
  absorbance and reflectivity of the material can be modulated by
  injection or extraction of electrons and ions, giving excellent
  control over and tunability to the optical properties\cite{elc1,elc2}.
  
  In addition to strongly affecting the color of the material, the
  incorporation of electron donating ions also has a strong effect on
  the structure.  For example, complex structural behaviour is
  observed on doping with electron donors, such as Na, Li or H.  Such
  dopants occupy the vacant perovskite A site and therefore they can
  be easily introduced over a wide range of concentrations, from trace
  concentrations in the so-called $\epsilon$ phase up to the limiting
  case of the tungsten bronze, NaWO$_3$. In NaWO$_3$, there is one
  donor per formula unit, and an aristotypic perovskite structure is
  adopted\cite{Kohlstrung}.  Its structure is the same as that of
  ReO$_3$, with both solids retaining the ideal cubic perovskite
  structure at all temperatures. Indeed, assuming that in NaWO$_3$ the
  valence electron of Na is donated into the {\em 5d}-type
  conduction/antibonding band of WO$_3$, the two materials are
  isoelectronic.  In contrast, WO$_3$ displays both off-centering of
  the W ion from its ideal centrosymmetric position, and
  Glazer\cite{Glazer72}-type tilting transitions. Studies on
  intermediate concentrations of Na$_x$WO$_3$ \cite{Clarke77} suggest
  a complex phenomenology, involving a consistent reduction in the
  degree of polyhedral tilting with increasing dopant
  concentration. Notably, Clarke observed that the room-temperature
  structures of Na$_x$WO$_3$ ($0.62 < x < 0.94$) are slightly
  distorted from the high-symmetry $Pm\bar{3}m$ aristotype, and
  proposed a preliminary phase diagram to explain the observed
  diffraction data\cite{Clarke77}.

  There have been several prior computational studies on WO$_3$,
  carried out at various levels of theory. Here we summarize the
  results of the density functional theory (DFT) studies from the
  literature. Perhaps the most relevant for our study is the work by
  Cor\`{a} et al.\cite{CoraDFT, Stachiotti} on the electronic
  structure of cubic WO$_3$, ReO$_3$ and NaWO$_3$. Using the full
  potential linear muffin tin orbital method, with the local density
  approximation, they showed that the band structures of the three
  compounds are very similar, with the extra electron in ReO$_3$ and
  NaWO$_3$ occupying the antibonding conduction band and decreasing
  the metal-oxygen bonding strength. They also showed that
  displacement of the transition metal from its centrosymmetric
  position towards an oxygen ion causes increased hybridization of the
  transition metal $5d$ $t_{2g}$ orbitals at the bottom of the
  conduction band with the O $2p$ orbtials at the top of the valence
  band. This in turn lowers the energy of the top of the valence band,
  and raises that of the bottom of the conduction band, explaining why
  metallic ReO$_3$ and NaWO$_3$ remain cubic, whereas ``$d^0$''
  WO$_3$\cite{hill} has an off-centering distortion.  Hjelm et
  al.\cite{Hjelm96} used the same method to establish that, in
  LiWO$_3$ and NaWO$_3$, rigid band filling of the WO$_3$ conduction
  band occurs, whereas in HWO$_3$ the hydrogens form hydroxide units
  with the oxygen atoms and change the electronic structure.  de Wijs
  et al\cite{deWijs} calculated the electronic properties of the
  various experimental structural phases of WO$_3$ using a plane wave
  ultra-soft pseudopotential implementation of DFT within both the
  generalized gradient and local density approximations. They found
  that increases in pressure are readily accomodated by tilting of the
  octahedra, explaining the small experimental bulk modulus. Also,
  they observed large W displacements, accompanied by strong
  rehybridization and changes in the electronic band gaps.  This
  latter observation is consistent with the anomalously large Born
  effective charges calculated for cubic WO$_3$ by Detraux et
  al.\cite{Detraux}.

  The work presented here builds on these earlier theoretical studies
  by treating {\it fractional} doping effects (i.e., the effect of
  electron concentrations between 0 and 1 per W ion) in a fully
  self-consistent manner.  Our computational approach allows the
  isolation of the effects of additional valence electrons from other
  factors, such as the presence of donor cations and structural
  disorder at intermediate concentrations. Indeed, it has already been
  shown that simple model calculations including only electronic
  effects can reproduce some aspects of the observed structural
  behavior \cite{CoraDFT}. Our main result is that electronic effects
  are able to account fully for the experimentally observed structural
  phase transitions.

  \section{Methodology}
  
  Calculations were performed using the {\sc Siesta} implementation
  \cite{Ordejon95,SIESTA}, of density-functional
  theory\cite{HohenbergKohn} (DFT), within the
  Perdew-Zunger\cite{Perdew} parametrization of the local-density
  approximation\cite{KohnSham}.  Core electrons were replaced by
  norm-conserving Trouiller--Martins\cite{Trouiller} pseudopotentials,
  factorised as prescribed by Kleinman and
  Bylander.\cite{KleinmanBylander} The valence electrons were taken to
  be $2s^22p^4$ for O, and $6s^25d^4$ for W.  Wave-functions were
  expanded in a basis of numerical atomic orbitals of finite
  support\cite{Anglada02}. A double-$\zeta$ polarised (DZP) basis set
  was used, whereby two basis functions per valence atomic orbital are
  included, plus extra shells of higher angular momentum to allow
  polarization.  This size of basis has been found, in prior studies,
  to give results of similar quality to typical plane-wave basis sets
  in other major codes. Details for the pseudopotentials and the basis
  sets are given in Tables~\ref{radii} and~\ref{pseudos}. 

\begin{table}
  \begin{ruledtabular}
  \begin{tabular}{ccccc}
	{\bf Species} & {\em n} & {\em l} & $r(\zeta_1)$ & $r(\zeta_2)$ \\
	\colrule
	W & 6 & 0 & 6.50 & 5.00 \\
	& 5 & 2 & 6.49 & 3.75 \\
        & 6 & 1$^*$ & 6.49 & n/a \\
	O & 2 & 0 & 3.31 & 2.51 \\
	& 2 & 1 & 3.94 & 2.58 \\
	& 3 & 2$^*$ & 3.94 & n/a \\
  \end{tabular}
  \end{ruledtabular}
  \caption{Cutoff radii, $r({\zeta})$, for the basis sets corresponding to
    W and O species; as the basis set is double--$\zeta$ polarized, there
    are two basis functions (and therefore radii) per hypothetical
    atomic orbital, as specified by quantum numbers $n$ and $l$. All
    radii are given in bohr. The starred entries are polarization orbitals.}
  \label{radii}
\end{table}

\begin{table}
  \begin{ruledtabular}
  \begin{tabular}{ccccc}
	{\bf Species} & $s$ & $p$ & $d$ & $f$ \\
	\colrule
	W $(5d^46s^2)$ & 2.85 & 3.03 & 2.25 & 2.25 \\
	O $(2s^22p^4)$ & 1.15 & 1.15 & 1.15 & 1.15 \\
  \end{tabular}
  \end{ruledtabular}
  \caption{Confinement radii for each of the $s$, $p$, $d$, and $f$
    channels of our W and O pseudopotentials. All radii are given in
    bohr. The W pseudopotential used also included a partial-core
    correction, with a radius of 1.30 bohr, according to the scheme of
    Louie et al\cite{LouiePCC}.}
  \label{pseudos}
\end{table}

  For $k$-point sampling, a cutoff of 10 \AA\/ was used\cite{Moreno}; this
  gave 24 independent $k$-points in the first Brillouin zone of the
  low-temperature monoclinic phase, corresponding to a
  4$\times$3$\times$3 mesh. This is equivalent to 6 independent
  $k$-points within the first Brillouin zone of a 64-atom supercell of
  the same structure, corresponding to a 2$\times$2$\times$3 mesh (if
  one were to neglect degeneracy). Both $k$-point meshes were
  generated by the method of Monkhorst and Pack\cite{MonkhorstPack}.
  The fineness of the real-space grid used for numeric integration was
  set to correspond to an energy cutoff of 200 Ry. A grid-cell
  sampling\cite{SIESTA} of four points (arranged on a face-centered
  cubic lattice) was used to reduce the space inhomogeneity introduced
  by the finite grid.  All calculations were performed using
  variable-cell relaxation, with the convergence criteria set to
  correspond to a maximum residual stress of 0.01 GPa, and maximum
  residual force component of 0.04 eV/{\AA}.

  In order to test the quality of the pseudopotential and basis set
  used, we first performed calculations on the same phases as investigated
  by de Wijs et al.  Results of comparable quality to theirs were
  obtained (see table \ref{latparams}), although our results tend to
  give the expected underestimation of the experimental lattice
  parameters whereas theirs, even in the LDA, tend to overestimate.
  
  In order to investigate the effect of additional dopant charge on
  the bonding character and structure of WO$_3$, a 2$\times$2$\times$1
  supercell of the low-temperature monoclinic ($\epsilon$) phase was
  used (the aforementioned 64-atom supercell). This contains sixteen W
  centers, and can be considered to be a $2\sqrt 2 \times 2\sqrt 2
  \times 2$ supercell of the aristotypic perovskite structure. As
  such, it is sufficiently large to encompass all ground-state
  symmetries experimentally observed in pure WO$_3$. Additional charge
  was added by the injection of extra electrons into the system, with
  charge neutrality over space being ensured by a corresponding
  homogeneous positively-charged background.
  
  \begin{table}
    \caption{Lattice parameters, experimental, from our simulations,
	and from the work of de Wijs\cite{deWijs} for (hypothetical)
	cubic, low temperature monoclinic, room temperature monoclinic,
        high-temperature tetragonal and room temperature triclinic
	phases of
	WO$_3$. All distances are given in \AA. Note
	that de Wijs' calculations were performed under the
	local-density approximation for monoclinic phases, and under
	the generalized gradient approximation for triclinic and
	tetragonal phases.}
    \begin{ruledtabular}
    \begin{tabular}{llcccccc}
      {\bf Phase} & & a & b & c & $\alpha$ & $\beta$ & $\gamma$ \\
      \colrule
      Cubic \\
      & This work & 3.81 & 3.81 & 3.81 & 90.0 & 90.0 & 90.0 \\
      Monoclinic & LT \\
      & This work & 5.15 & 5.05 & 7.63 & 90.0 & 92.6 & 90.0 \\
      & de Wijs & 5.34 & 5.31 & 7.77 & 90.0 & 90.6 & 90.0 \\
      & Experiment\cite{Salje97}& 5.28 & 5.15 & 7.66 & 90.0 & 91.8 & 90.0 \\ 
      Monoclinic & RT \\
      & This work & 7.30 & 7.49 & 7.32 & 89.8 & 90.1 & 90.0\\
      & de Wijs & 7.37 & 7.46 & 7.64 & 90.0 & 90.6 & 90.0 \\
      & Experiment\cite{Loopstra}& 7.31 & 7.54 & 7.66 & 90.0 & 91.8 & 90.0\\
      Tetragonal  \\
      & This work & 5.31 & 5.31 & 3.91 & 90.0 & 90.0 & 90.0 \\ 
      & de Wijs & 5.36 & 5.36 & 3.98 & 90.0 & 90.0 & 90.0\\
      & Experiment\cite{SaljeT}& 5.27 & 5.27 & 3.92 & 90.0 & 90.0 & 90.0 \\ 
      Triclinic & RT \\
      & This work & 7.18 & 7.36 & 7.63 & 88.0 & 90.5 & 90.4 \\
      & de Wijs & 7.54 & 7.64 & 7.84 & 89.7 & 90.2 & 90.2 \\
      & Experiment\cite{Tanisaki}& 7.30 & 7.52 & 7.69 & 88.8 & 90.9 & 91.0 \\ 
    \end{tabular}
    \end{ruledtabular}
    \label{latparams}
  \end{table}
  \section{Calculated phase diagram}

  \begin{figure}
    \includegraphics[width=3in, angle=0]{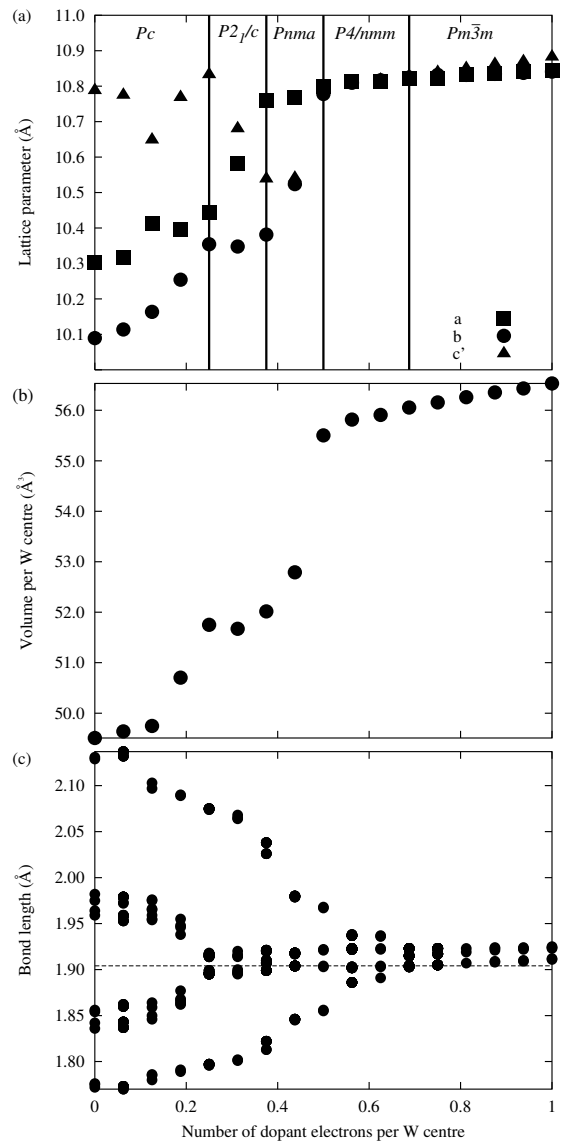}
    \caption{(a) Lattice parameters (for normalized cubic supercell
      with respect to the $\epsilon$ phase) of WO$_3$; (b) Volumes of
      2$\times$2$\times$2 supercell of WO$_3$; (c) Bond lengths within
      WO$_6$ octahedra, all versus electron dopant concentration.}
    \label{Lattice Parameters}
  \end{figure}

  We performed calculations for numbers of dopant electrons between 1
  and 16 per supercell, (corresponding to electron concentrations in
  the range $0 < x < 1$ per W center) in single-electron steps. Full,
  variable-cell, structural relaxations were performed for each doping
  level. A series of five phases was observed over the doping
  range. Here we describe the five structures in turn, starting with
  the highest doping level. We refer to figure~\ref{Lattice
  Parameters}, which shows (a) our calculated lattice parameters ({\em
  c'} is the {\em c} lattice parameter multiplied by $\sqrt{2}$ in
  order to normalize it with respect to the {\em a} and {\em b}
  parameters), (b) our calculated unit cell volumes, and (c) our W-O
  bond lengths. Our suggested phase boundaries are also shown in
  figure~\ref{Lattice Parameters} (a) as solid vertical lines.

  At maximal doping, we obtain the perovskite aristotypic (i.e. the
  highest symmetry phase in the series: in the case of perovskites,
  primitive cubic, $a=b=c$, one formula unit per unit cell) structure,
  with space group $Pm\bar{3}m$. The three lattice parameters are
  equal, as are the W-O bond lengths. 

\begin{figure}
    \includegraphics[height=2in]{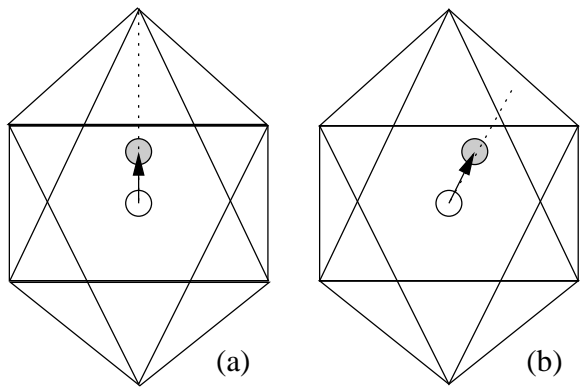}
    \caption{Offcentering in WO$_6$ octahedra; (a) along [001], (b)
    along both [001] and [110]-type directions}
    \label{offcent}
  \end{figure}

  With decreasing dopant
  concentration the unit cell volume decreases slightly, until at $x
  \approx \frac{11}{16}$, we find a symmetry-breaking transition to an
  antiferroelectric tetragonal phase with $P4/nmm$ symmetry. The W atom 
  moves off the center
  of its O$_6$ octahedron in the [001] direction [as shown in figure
  \ref{offcent} (a)], resulting in different W-O bond lengths and a
  lowering of the space group of the system. Interestingly, all three
  lattice parameters continue to have the same effective length with
  respect to our supercell (within the accuracy of our computations)
  in spite of the inequivalence of one of them due to the reduction in
  symmetry. This can be ascribed to the fact that the oxygen framework
  distorts very little in these transitions, and it is the oxygen
  framework that is largely responsible for the volume and shape of
  the unit cell. The antiferroelectric tetragonal phase
  persists down to a dopant concentration of
  approximately $\frac{1}{2}$ electron per W, with a gradual increase
  in off-centering with decreasing dopant concentration. By
  $x=\frac{1}{2}$, the off-centering of the W along the [001] direction
  has increased to around 0.08 \AA.
  
  The antiferroelectric tetragonal phase becomes unstable under
  $\frac{1}{2}$ electrons per W, and an orthorhombic phase is
  stabilised. There is also a clear discontinuity in the volume per
  unit cell at this doping concentration. This arises from a tilt of
  the WO$_6$ octahedra towards $[110]$-type directions which in turn causes
  the loss of the four-fold rotation in the $c$ direction, lowering
  the space group to $Pnma$. Initially, the $c'$ and $a$ lattice
  parameters (within our pseudocubic supercell), despite being
  symmetrically inequivalent, remain essentially equal in length; this
  can be ascribed to the absence of distortion of the WO$_6$
  octahedra, which originally tilt (around an axis parallel to $b$) as
  a rigid unit without substantial deformation. These parameters
  become unequal at x $\approx\frac{7}{16}$; however, we do not find a
  lowering of the space group caused by this.

  Below $x\approx \frac{3}{8}$, a monoclinic $P2_1/c$ phase (similar to that
  identified by de Wijs\cite{deWijs}), related to the low-temperature
  $\epsilon$ phase in WO$_3$, is stable. The phase boundary from the
  orthorhombic to monoclinic phase is characterized by an increase in
  the W-O displacement along [001] (seen in the increased bond
  splitting in figure~\ref{Lattice Parameters} (c)) as well as a
  marked rotation around $z$ (loosely speaking in the $x-y$ plane).
  The Glazer tilt is expressible as $a^-b^-c^-$.  The monoclinic
  symmetry persists through the remainder of the phase diagram;
  however an additional phase boundary exists at $x\approx
  \frac{1}{4}$.  Here an additional splitting in the W-O bond lengths
  indicates offcentering of the W atom in a [110]-type direction (as
  shown in figure~\ref{offcent} (b)) and there is also a second
  discontinuity in the cell volume. This corresponds to a loss of
  screw axes along b, and hence a further lowering of the space group
  to $Pc$, the $\epsilon$ phase.

\begin{figure}
    \includegraphics[width=3in, angle=0]{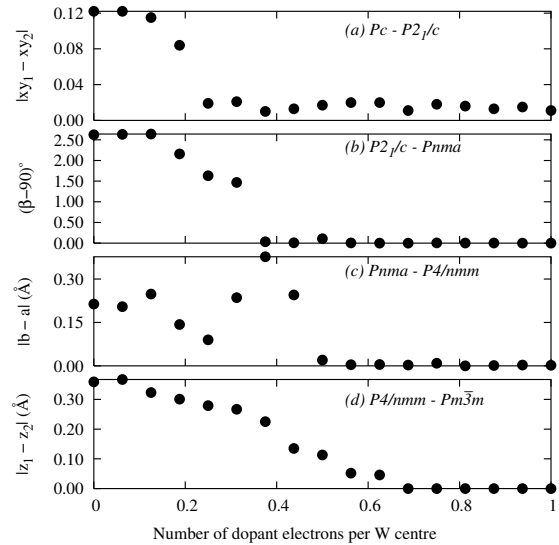}
    \caption{Order parameters for the doping-induced phase transitions
      in WO$_3$. (a) $Pc$ -- $P2_1/c$, in \AA, where $xy_1$
      and $xy_2$ are the inequivalent W-O bond lengths in the the
      $x-y$ plane; (b) $P2_1/c$ -- $Pnma$, in degrees; (c)
      $Pnma$ -- $P4/nmm$ (in \AA); (e) $P4/nmm$ -- $Pm\bar{3}m$ (in \AA),
      where $z_1$ and $z_2$ refer to the two W--O bonds aligned
      predominantly with the $z$ axis.}
    \label{OrderParameters}
  \end{figure}

  Order parameters can be defined for the transitions between each of
  these phases as shown in figure~\ref{OrderParameters}. The order
  parameter for the $Pc$ to $P2_1/c$ transition
  (figure~\ref{OrderParameters} (a) )is the difference between the W-O
  bondlengths in the $x-y$ plane ($xy_1$ and $xy_2$). The order
  parameter for the $P2_1/c$ to $Pnma$ transition is, as expected, the
  deviation of the $\beta$ angle from 90 $^{\circ}$; this goes sharply
  to zero at $x \approx \frac{3}{8}$ (figure ~\ref{OrderParameters}
  (b)).  The order parameter for the $Pnma$ to $P4/nmm$ phase is the
  length difference between the two remaining unequal lattice
  constants, $|b-a|$ (figure~\ref{OrderParameters} (c)).  Finally, the
  ideal cubic $Pm\bar{3}m$ perovskite structure is reached when the W
  moves to its centrosymmetric position, and the order parameter for
  the tetragonal to the cubic phase is the magnitude of the W
  off-center displacement along [001] (figure~\ref{OrderParameters}
  (d)).  As such, all the phase boundaries are well-defined, although
  further work could be undertaken to locate them with greater
  precision. In the next Section we discuss the effects giving rise to
  this phase structure in greater depth.

\section{Discussion}

  \begin{figure}
    \includegraphics[height=2.5in, angle=0]{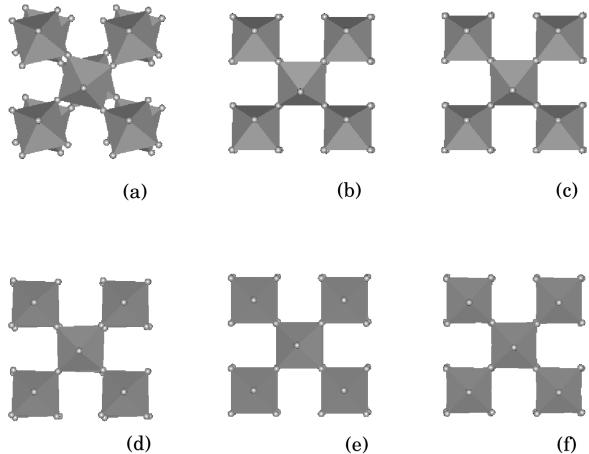}
    \caption{Polyhedral tilts in WO3, as varying with dopant charge.
The charges in each case are, in electrons per W atom, are; (a) 3/16,
(b) 5/16, (c) 7/16, (d) 1/2, (e) 5/8, (f) 1; these correspond to,
respectively, the monoclinic $Pc$ phase; the $P2_1/c$ phase,
approaching the $Pnma$ boundary; the middle of the orthorhombic phase;
the $Pnma$ -- $P4/nmm$ boundary; within the stability range of the
tetragonal phase; and the aristotype.  It is clear that the
distortions between some phases are extremely small. This poses
difficulties in ascertaining exactly where they are located in
composition space.  }
    \label{tilts}
  \end{figure}

  We see that there are two types of structural distortion occurring
  in WO$_3$; rotation of the (almost rigid) octahedra, and
  displacement of the W ion from the center of its octahedron.  The
  occurrence or absence of W offcentering is known to be determined by
  a balance between electronic Coulomb repulsions (which are minimized
  for the centrosymmetric structure) and additional bonding
  considerations which might stabilize the non-centrosymmetric
  phase\cite{hill}. In the case of WO$_3$, off-centering of the W ion
  results in additional hybridizations that lower the energy of the O
  $2p$-like valence band compared with that of the centered structure,
  and raise the energy of the predominantly W $5d$ $t_{2g}$ conduction
  band\cite{CoraDFT}. As the doping level is increased, and more
  electron density is added to the conduction band, there is no
  energetic advantage to these additional hybridizations, and the W
  ion moves back to its centrosymmetric position.

  The fact that the small off-centering in the [110] direction 
  is quenched first implies that the antibonding orbitals
  corresponding to this covalent interaction are low-lying in the
  conduction band and therefore filled first.  The offcentering in the
  {\em z} direction persists to higher doping concentrations and is only 
  quenched out at $\approx \frac{3}{4}$ e$^-$ per center.
  This 75\% concentration is clearly less than the value of
  0.98 e$^-$ suggested by Cor\`{a} et al.'s analysis of one-electron
  energies for the displacement of Re along [100] in
  ReO$_3$\cite{CoraDFT}; this accords with their suggestion, however,
  that crystal--field effects would favour cubic structures, and thus
  any analysis under a rigid-band approximation (or similar) would
  overestimate the degree of doping necessary to cause the onset of
  cubicity.

  There is strong evidence of a relationship between the polyhedral
  rotation and offcentering mechanisms. In figure \ref{tilts} we show
  sketches of the structures at different doping concentrations to
  illustrate the polyhedral tilting.  At $\frac{3}{16}$ doping (figure
  \ref{tilts} (a)) we see that the rotations are large around both
  axes. However in figure \ref{tilts} (b) (by which the offcentering
  in the [110] direction has been essentially quenched ($x\approx
  \frac{5}{16}$)) we see that the rotation of polyhedra in that plane
  also disappears, whilst rotation of polyhedra between connections
  along the $z$ axis persists, as does the offcentering in that
  direction. This persists, with reducing magnitude, to x$\approx
  \frac{7}{16}$ (figure \ref{tilts} (c)), but by 50\% doping, (figure
  \ref{tilts} (d)) any remaining tilt up the $z$ axis (out of the page) is
  extremely small. However, at this concentration, offcentering along
  [001] persists. Notice also that the majority of volume change in
  the unit cell, which in perovskites is caused by changing of the
  rotation angles, has also occurred by this doping level. Finally,
  any tilt systems in figures \ref{tilts} (e) and (f), corresponding to
  concentrations with centered W ions, are too small to observe in the
  figure.

  It should be noted that the very small [100]-offcentering - less than
  0.01 \AA\ in magnitude - which persists at high doping concentration
  does not appear to be a real effect. In fact, both this and the
  small Jahn-Teller distortion at 100\% doping appear to be numerical
  artifacts related to the real-space mesh cutoff used: when the mesh
  is increased to 500 Ry (from 200 Ry), both effects disappear. As it
  is computationally expensive, one cannot justify this level of
  convergence for all calculations, particularly given that the order
  parameters delineating the phase boundaries in the system are
  well-defined at the present level of accuracy. Even so, it is clear
  that the system is very near the limits where Jahn-Teller distortion
  and/or offcentering of the W atom become thermodynamically stable.
  As expected by analogy with ReO$_3$, WO$_3$ doped with 1 electron is
  entirely cubic, with no Jahn-Teller distortion or W
  off-centering. Since the transition metal, in both cases, has a
  formal $d^1$ electron configuration (and $d^1$ ions in octahedrally
  coordinated complexes do not have a second-order Jahn-Teller
  distortion) the centrosymmetricity is easy to explain. The lack of a
  true Jahn-Teller distortion (which is predicted for such a $d^1$ ion
  in octahedral coordination) is more subtle however, and is
  determined by a competition between energy lowering through
  hybridization (strongest in the cubic structure) and energy lowering
  through gap formation during the distortion. In bulk ReO$_3$, the
  broad bands and low density of states at the Fermi energy result in
  the cubic phase having the lowest energy. In contrast, in a
  molecular complex the bands are infinitely narrow and the Fermi
  energy density of states is infinitely large (favouring Jahn-Teller
  distortion).

  \begin{figure}
    \includegraphics[height=3.5in, angle=270]{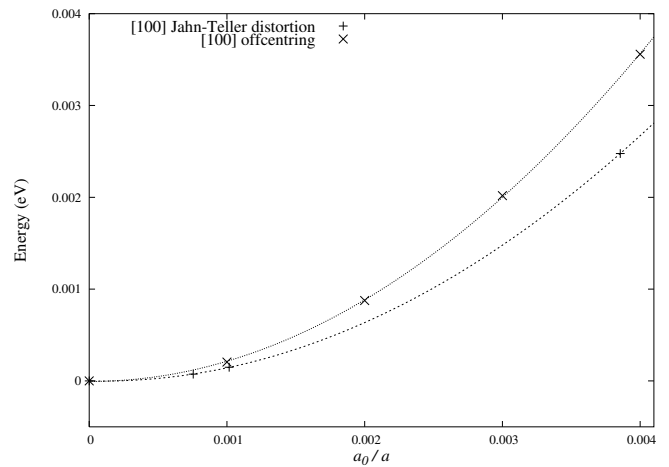}
    \caption{Increase in energy as a function of fractional change in W-O
bondlength for both W off-centering and Jahn-Teller distortion
along a cubic axis.} 
    \label{JT_vs_offcenter}
  \end{figure}
  
  In figure ~\ref{JT_vs_offcenter} we plot our calculated change in
  energy, as a function of the amount of {\it both} Jahn-Teller
  distortion and transition metal off-centering, for WO$_3$ doped with
  one electron per W center. In both cases a constant volume is
  maintained. The $x$ axis shows the fractional change in bond length;
  for the Jahn-Teller distorted structure both W-O bondlengths
  increase by this amount, and for the off-centered structure one
  increases and the other decreases. Note that, by this measure, the
  structure is stiffer to Jahn-Teller distortion than it is to
  off-centering. Also note that our curvature for off-centering is
  very close to that of Cora et al.'s\cite{CoraDFT} for ReO$_3$ (which
  in turn is around half that for NaWO$_3$).
  
  \begin{figure}
    \includegraphics[width=3.0in]{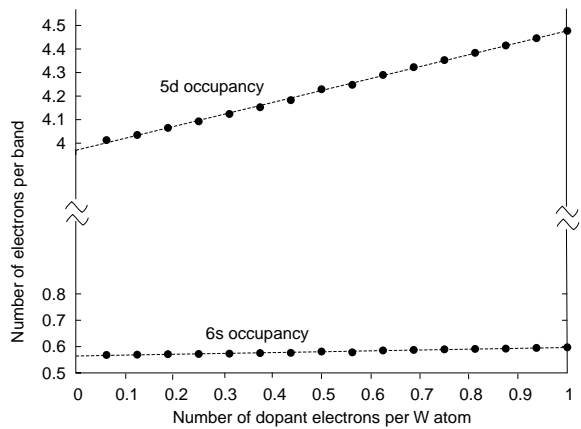}
    \caption{Mean occupancy per W atom of $5d$ and $6s$ orbitals 
    versus dopant charge.}
    \label{dbandoccupancy}
  \end{figure}

  A Mulliken analysis of the occupancy of the three families of bands
  due to W -- $6s$, $5d_{e_g}$, and $5d_{t_{2g}}$ -- suggests that the
  occupancy of the $5d$ band is far from $d^0$ in undoped WO$_3$, and
  far from $d^1$ in the fully doped system. The occupancy of the
  $d$-band is closer to four in undoped $\epsilon$--WO$_3$ (see
  figure \ref{dbandoccupancy}) and increases by approximately 0.5
  e$^-$ in the fully doped system. The remaining half of an electron,
  unsurprisingly, fills the antibonding bands formed by
  O($2p$)--W($5d$) overlap.  Note that the mean occupancy of the $6s$
  orbital, which is $\approx$ 0.58 electrons per orbital in the
  undoped system, hardly increases on addition of electrons.  This
  implies a typical W valence population of between 4.5 and 5
  electrons, in contrast to the idealised purely-ionic picture of a
  bare W$^{6+}$ ion (with no valence electrons); this is yet further
  evidence of a system remarkably dominated by covalent W--O
  interactions.  It is clear from further analysis that above 50\%
  doping, dopant charge preferentially fills the $5d_{xz}$ and
  $5d_{yz}$ bands; the other bands experience only a small change in
  total occupancy. This accords well with the suggestion that
  bonding--antibonding splitting occurs through O($2p$)-W($5d$)
  $\pi$-like overlap; the absence of change in $xy$ bonding being
  demonstrated by the absence in change of occupancy of the $5d_{xy}$
  band. However, given the tilting transitions and highly-deformed
  structures, there is great difficulty in assigning the occupancy of
  $5d$ orbitals; therefore, further studies of the influence of
  band formation and orbital occupation on both first- and
  second-order Jahn-Teller distortions are ongoing.

  In terms of reconciling our calculations with previous experimental
  data\cite{Clarke77}, one must be somewhat cautious. Clarke analyzed
  his results under the presumption that there would not be
  significant offcentering of the W ion within WO$_6$ octahedra, which
  we believe to be incorrect: nevertheless, we believe that our
  tetragonal phase is consistent with collected data for his proposed
  cubic phase at high doping. Furthermore, our calculations
  deliberately neglect the effects of defects, chemical disorder, and
  other imperfections in the lattice.

  Finally, we mention that the original motivation of our study was to
  investigate possible self-trapping behaviour of charge in this
  material. Excess electrons in the $\epsilon$ (low-temperature
  monoclinic) phase have been shown experimentally to self-trap,
  forming polarons\cite{Salje97}. We found no localisation of charge
  or deformation (polaron formation) in any of our calculations. This
  null result is of some interest, but should not be taken to mean
  that polarons do not form in this material; it is arguable that
  either the supercell used is too small to observe polaron
  localisation, or that the LDA will underestimate the binding energy
  of a polaron (thus causing it not to be a stable state of the
  system): further study is needed in this area.

  \section{Conclusions}
  
  The main conclusion that we draw from our calculations is that the
  experimentally observed structural distortions induced in
  Na$_x$WO$_3$--type bronzes by increasing doping are predominantly
  electronic in nature. By using our methodology of adding electrons
  to WO$_3$ without also adding Na atoms, we have removed any possible
  structural/disorder effect caused by the Na$^+$ cations on the A
  sites within the structure. Therefore, given that our calculations
  reproduce the experimental observed sequence of symmetries upon
  doping ((i) Monoclinic ($Pc$ and $P2_1/c$); (ii) Orthorhombic
  ($Pnma$); (iii) Tetragonal ($P4/nmm$); (iv) cubic aristotype
  ($Pm\bar{3}m$)) we conclude that the effect of the Na$^{+}$ cations
  on the structure must be small. Indeed, this is not unexpected.
  Following the classification of Robin and Day\cite{Robin}, as
  mentioned by Bersuker\cite{Bersuker}, NaWO$_3$ is a good example of
  pure electronic doping; there is complete transfer of the donated Na
  electron over to the WO$_3$ sublattice since the 3{\em s} band of Na
  lies well above the 5{\em d}-antibonding band from W.
  
  Finally, we propose that these materials present interesting
  opportunities for future experimental and theoretical study, given
  the degree of structural control that can be gained from doping or
  substitution.  In particular, the ``quenching out'' of polyhedral
  rotation (and hence the opening of channels within the structure)
  may have effects on diffusion and ion intercalation in these
  structures.
  
  \begin{acknowledgments}
  
  We would like to thank colleagues within the Mineral Physics group
  for technical assistance, and Dr. Martin Dove, Dr. William Lee,
  Dr. Simon Redfern, Prof. Ekhard Salje, and Prof. James Scott for
  fruitful discussions.  Nicola Spaldin thanks the Earth Sciences
  Department at Cambridge University for their hospitality during her
  recent sabbatical leave, and the U.S. National Science Foundation
  for financial support under grant number DMR-0312407.  Andrew
  Walkingshaw wishes to acknowledge financial support from the
  Engineering and Physical Sciences Research Council, United Kingdom.

  \end{acknowledgments}

\bibliographystyle{apsrev}
\bibliography{adw}
\end{document}